\documentclass{article}
\usepackage{arxiv}
\usepackage[utf8]{inputenc} 
\usepackage[T1]{fontenc}    
\usepackage{hyperref}       
\usepackage{url}            
\usepackage{booktabs}       
\usepackage{amsfonts}       
\usepackage{nicefrac}       
\usepackage{microtype}      
\usepackage{lipsum}		    
\usepackage{graphicx}
\usepackage{natbib}
\usepackage{doi}
\usepackage{amsmath,amssymb,amsthm}
\graphicspath{{figures/}}

\title{A study of CNN capacity applied to Left Ventricle segmentation in cardiac MRI\thanks{This preprint has not undergone peer review (when
applicable) or any post-submission improvements or corrections. The Version of Record of this article is published in SN Computer Science, and is available online at \url{https://doi.org/10.1007/s42979-021-00897-x}.}}

\renewcommand{\cite}[1]{\citep{#1}}
\newcommand{\orcid}[2]{\href{https://orcid.org/#1}{\includegraphics[scale=0.07]{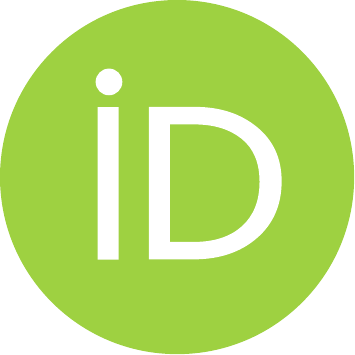}\,#2}}

\author{\orcid{0000-0001-6585-0354}{Marcelo Arruda}\thanks{Corresponding author email: \texttt{marcelo.arruda@hc.fm.usp.br}},
\orcid{0000-0002-7818-6103}{Daniel Lima},
\orcid{0000-0001-5464-1792}{José Krieger},
\orcid{0000-0003-0964-6222}{Marco Gutierrez} \\
	Biomedical Informatics Lab\\
	Heart Institute (InCor/HC.FMUSP)\\
	Avenida Doutor Eneas Carvalho de Aguiar 44, Cerqueira César, 05403-904, Sao Paulo, SP, BR}

\date{May 10, 2021}

\renewcommand{\headeright}{{arXiv} Preprint}
\renewcommand{\undertitle}{{arXiv} Preprint}
\renewcommand{\shorttitle}{Study of CNN capacity for LV segmentation in MRI}

\hypersetup{
pdftitle={A study of CNN capacity applied to Left Ventricle segmentation in cardiac MRI},
pdfsubject={q-bio.NC, q-bio.QM},
pdfauthor={Marcelo A.~F.~Toledo, Daniel M.~Lima, José E.~Krieger, Marco A.~Gutierrez},
pdfkeywords={deep learning, model selection, magnetic resonance imaging},
}

\begin{document}
\maketitle

\begin{abstract}
CNN (Convolutional Neural Network) models have been successfully used for segmentation of the left ventricle (LV) in cardiac MRI (Magnetic Resonance Imaging), providing clinical measurements. In practice, two questions arise with deployment of CNNs: 1) when is it better to use a shallow model instead of a deeper one? 2) how the size of a dataset might change the network performance? We propose a framework to answer them, by experimenting with deep and shallow versions of three U-Net families, trained from scratch in six subsets varying from 100 to 10,000 images, different network sizes, learning rates and regularization values. 1620 models were evaluated using 5-fold cross-validation by loss and DICE. The results indicate that: sample size affects performance more than architecture or hyper-parameters; in small samples the performance is more sensitive to hyper-parameters than architecture; the performance difference between shallow and deeper networks is not the same across families.
\end{abstract}


\section{Introduction}
\begin{figure}[htb]
    \centering
    \fbox{\includegraphics[height=.23\columnwidth]{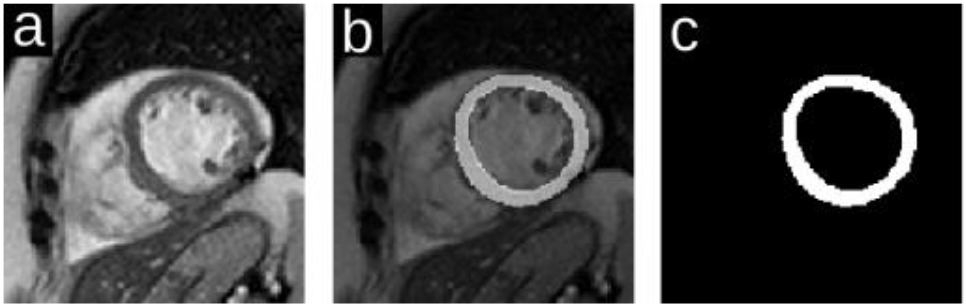}}
    ~
    \fbox{\includegraphics[height=.23\columnwidth]{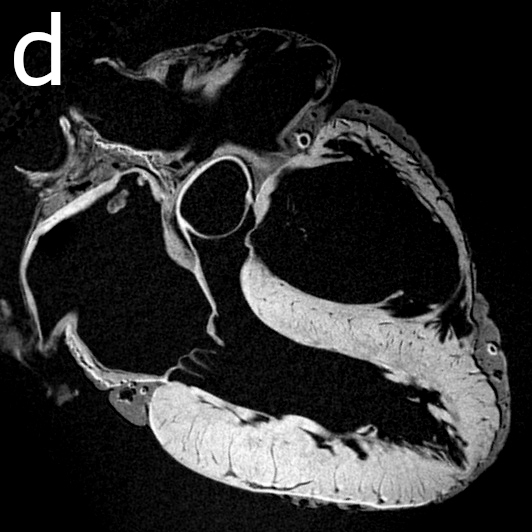}}
    \caption{Cardiac MRI images are acquired in two different axis: short-axis (a-c) and long-axis (d). In (a) we see a cardiac short-axis MRI of heart, centred in the heart, and cropped to frame the right and left ventricles; in (b) the left ventricle myocardium is highlighted, and (c) only shows the left ventricle myocardium segmentation mask. Images (a-c) are reproduced from the dataset analysed in \cite{nikolov2018deep} and (d) is reproduced from the Atlas of Human  Cardiac Anatomy website \cite{spencer2013}.}
    \label{fig:lv}
\end{figure}

The use of deep learning has been growing and becoming very relevant for some types of medical image analysis. Deep learning includes CNN (Convolutional Neural Network) models, which have been successfully used for segmentation of the left ventricle (LV) in cardiac MRI (Magnetic Resonance Imaging) by \citet{chen2020deep}. The myocardium segmentation (Figure \ref{fig:lv}) is the basis for other steps that provide measurements of ventricular function and myocardial viability (such as ejection fraction and ventricular mass), and is used for diagnosis of heart diseases, e.g. coronary artery disease \cite{faridah2013state}. One of the most successful CNN architectures for medical image segmentation is the U-Net \cite{ronneberger2015u}. It has been successfully applied in several medical imaging analyses such as pancreas detection \cite{oktay2018attention}  and prostate segmentation \cite{ghavami2019automatic}. Figure \ref{fig:lv} shows an example of LV segmentation in cardiac MRI.

One of the limitations of deep learning approaches in many medical applications is the need for large labeled datasets, which are not always available. Creating labelled datasets requires clinical specialists for visual inspection and manual marking of images, which is costly and takes significant time. By considering this limitation, the U-Net architecture was designed for fast convergence and sufficiently accurate segmentation in relatively small datasets \cite{oktay2018attention}. U-Net is a CNN architecture with many possible variations such as different neural blocks, block width and network depth, so it is possible to balance the performance according to the images under study.

However, with the deployment of CNNs in medical practice, another two questions arise: 1) when is it better to use a small and shallow model instead of a larger and deeper one? 2) how the number of samples in a dataset might change the network performance? These questions motivate our work, which provides a framework of analysing the CNN capacity and dataset sizes for medical applications, and demonstrate its applicability to LV segmentation in cardiac MRI.

\subsection*{Our Contributions}

In order to grasp how dataset size affects U-Net performance in segmentation, this paper describe a framework where we train 18 different U-Nets of different sizes (number of network parameters and depth) and families: VGG \cite{simonyan2014very}, ResNet \cite{he2016deep} and EfficientNet \cite{tan2019efficientnet}. These U-Net variations were trained from scratch in six dataset sizes (suppose patients submitted to MRI study), and also with variations of hyper-parameters learning rate and L2 regularization. They were evaluated with 5-fold cross-validation by the loss and DICE index, and each fold was tested in a separate holdout set. We then analyse the variables and results in a factorial study using GAM (Generalized Additive Model), ANOVA (Analysis of Variance) and HSD (Honest Significant Difference) tests. In summary, our objectives in this paper are:

\begin{itemize}
 \item Observe the relationship between prediction error, dataset size and network size;
 \item Identify hyper-parameter effects on performance by linear models, ANOVA and HSD;
 \item Provide a framework to check when the choice for the network to be used in an application might change due to the number of images available in a training dataset.
\end{itemize}

This paper is organized as follows: Section \ref{sec:review} (Related work) reviews the base CNN architectures for image segmentation and related studies of neural network capacity; Section \ref{sec:method} (Materials and Methods) present the dataset preparation, the neural networks, hyper-parameters, linear models, ANOVA and HSD tests; Section \ref{sec:exps} (Experiments) present the experiments, metrics, results and discussion; and Section \ref{sec:conclusion} (Conclusion) summarizes our goals, the proposed framework, achievements and ideas for future investigations.

\section{Related work}
\label{sec:review}

In this section we review related literature in the two main aspects of our framework: biomedical image segmentation (\ref{sec:segment}), and how to evaluate neural network capacity (\ref{sec:capacity}).

\subsection{Biomedical Segmentation with U-Net and its backbones}
\label{sec:segment}

U-Net is an encoder-decoder CNN architecture for biomedical image segmentation, which encodes the input image in a sequence of four down-scaling layers (blocks), then up-scales the encoded tensors, concatenates them to the previous input and combines the results in the reversed order, forming a network that resembles an “U”. It is possible to change its backbone by replacing the original convolution blocks for well-known blocks from other CNNs, e.g. residual blocks \cite{zhao2017pyramid}, while keeping the U shape encoder decoder structure. The U-Net is trained by minimizing the cross-entropy between the class probabilities of each pixel in the output image and the segmentation mask given as class labels. The success of U-Net led to the proposal of backbone innovations, such as newer versions of the skip connections \cite{chaurasia2017linknet,zhou2018unet}. Other encoder-decoder architectures have also been proposed, some keeping the general U-net structure, and others with novel structures such as feature pyramids \cite{lin2017feature} and pooling pyramids \cite{zhao2017pyramid}. The standard U-Net backbone is based on VGG \cite{simonyan2014very}, but we extended our analysis by considering two more state-of-the-art backbones: ResNet \cite{he2016deep} and EfficientNet \cite{tan2019efficientnet}.

\subsection{Evaluating Neural Network Capacity}
\label{sec:capacity}

Given that there are several U-Net architectures available, making an appropriate choice is challenging and can impact the results of developing applications. These reflect in two critical aspects of medical studies planning: a) estimating the size of the sample needed for developing an effective solution (both in terms of images and patients) and b) selecting appropriate models, because different medical applications solve problems of different complexities, which thus require different network models of adequate capacity. However, there is no established method for calculating network capacity nor a consolidated understanding of how it relates to CNN depth and to the number of parameters \cite{bartlett2017spectrally, neyshabur2019towards}. Addressing these aspects in a clinical study is not trivial. While bigger networks might be prone to over-fit in small datasets, smaller networks might not have the required capacity for the challenge at hand. A further complication, at least for LV segmentation, is the fact that the relationship between U-Net, dataset sizes and generalization error is not well understood. Recent literature \cite{benkendorf2020effects,shahinfar2020many} have been investigating sample size issues, so in our approach we extend this investigation to hyper-parameters and network architecture. Next we present our methodology, then experiments, results and conclusion.

\section{Materials and Methods}
\label{sec:method}

In this section we present the dataset characteristics, preparation and splitting (Section \ref{sec:dataprep}, the U-Net CNN architecture and backbones based on VGG, ResNet and EfficientNet (Section \ref{sec:neural}),  hyper-parameter optimization (Section \ref{sec:hyper}), and the linear models and statistical tests used in the analysis (Section \ref{sec:linear}).

\subsection{Dataset preparation}
\label{sec:dataprep}

All of the datasets we created are sub-samples of the publicly available Left Ventricle Segmentation Challenge (LVSC) dataset \cite{nikolov2018deep}. The LVSC training set has cardiac cine-MRI pictures of 100 patients, with diverse number of slices (8–24), slice thickness (6–8mm), slice gaps (2–4mm), number of phases in the cardiac cycle (18–35), image sizes (138×192 to 512×512 pixels) and acquisition devices (GE, Phillips or Siemens MRI scanner systems with 1.5 or 3.0T). All images were loaded into float32 arrays, had the pixel values rescaled to [0,1], resized to 256×256 pixels with area and bi-cubic interpolation respectively for downscaling and up-scaling, and zero padding to maintain aspect ratio.

Due to anatomical and positional differences, patient id is a major source of variation, so we sub-sampled 100 images from each patient. In order to stratify samples in the whole cardiac volume, we selected an approximately equal number of images from each slice, regularly spaced throughout the cardiac cycle (images $\lfloor i \times phases \times slices / 100 \rfloor ~\mathrm{for}~ i ~\mathrm{in}~ [1, 100]$). This resulted in 10,000 images (100 patients × 100 images per patient). Then, an 80-20\% holdout group split was performed for development (training + validation) and test.
Afterwards, five smaller datasets were sub-sampled from the bigger dataset (10k), with 200, 500, 1000, 2500, and 5000 images. For each of those datasets, the development set was sub-sampled from the initial development set with 8000 images, while the test set was sub-sampled from the holdout test set with 2000 images. Both the holdout and the sub-sampling were performed by randomly selecting of patients, such that a given patient had either all of her/his images in one set only. In each execution the development set is again split in 5-folds of equal sizes for cross-validation: each fold is taken for validation using a model trained on the remaining four folds. Next we present the neural networks, hyper-parameters and linear models and statistical tests.

\subsection{Neural Networks}
\label{sec:neural}

U-Net \cite{ronneberger2015u} implements the downscaling path similarly to the VGG network \cite{simonyan2014very}, using a sequence of 2D convolutions, ReLU activations and max-poolings. We also tested U-Net with modern backbones based on ResNet \cite{he2016deep} and EfficientNet \cite{tan2019efficientnet}. ResNets divide the pathway in an input identity map path (a short-cut connection), and a residual path, that learns a residual function (which model the residuals from the inputs). By learning the residuals instead of the main signal, more convolution blocks can be stacked to build deeper neural networks without compromising their training. EfficientNets are based on mobile inverted bottleneck convolution blocks\cite{sandler2018mobilenet} with squeeze–excitation optimization\cite{hu2018squeeze} and a loss function that balances accuracy and speed, by adjusting network depth, width and resolution hyper-parameters \cite{tan2019efficientnet,tan2019mnasnet}.

To study network size we selected small and medium-large sizes for each backbone: EfficientNets B0 and B5; ResNets 18 and 50, with approximately the same numbers of parameters to the chosen EfficientNets; and VGG 16 and 19, the two most common VGG sizes in literature. EfficientNet B0 has 5.3 million parameters and achieved 76.3\% top-one accuracy in the ImageNet classification task, while B5 has 30 million parameters and achieved 83.3\% accuracy in that same task. This is a bigger gain for scaling up than for other families of networks \cite{tan2019efficientnet}. Another aspect that impacts performance and convergence of neural networks is the choice of hyper-parameters, presented next.

\subsection{Hyper-parameter optimization}
\label{sec:hyper}

Given that training strategies might differ regarding dataset and network size, we also chose to analyse some optimization parameters. However, as the time to compute all hyper-parameter space is combinatorial, only some values were selected. For each model and dataset combination, we tested different initial learning rates ($10^{-2}$, $10^{-3}$ and $10^{-4}$) and different L2 regularization values ($10^{-2}$, $10^{-4}$ and $10^{-6}$). The initial learning rate balances training speed and convergence stability. The regularization has a further implication in model capacity, as it balances accuracy and over-fitting by constricting network weights. In order to understand the effect of hyper-parameters in combination to dataset and network size, we analyse these variables with linear models, ANOVA and HSD tests of next section.

\subsection{Linear Models, ANOVA and HSD}
\label{sec:linear}

To understand the effect of CNN choices on the performance, we develop a linear regression model based on the dataset size, model architecture, model size, and hyper-parameters as input variables.
Linear regression models are statistical approaches to analyse the dependence of a response variable $Y$ on the linear combination of input variables $X = x_1, x_2, ..., x_n$. These models follow the general formula $Y = aX + b + \epsilon$, where $a$ corresponds to weights for each variable $x_i$, and $b$ is a bias term (intercept), and $\epsilon$ is the error of the model. The parameters $a$ and $b$ are commonly adjusted by ordinary least squares regression (OLS). ANOVA for regression models tests the statistical significance of input variables (groups), with respect to the sums of squares (SS) of regressed variables and observed data. The HSD test was developed by Tukey for pairwise comparison of multiple group means, and is used as a post hoc adjustment \cite{tabachnick2007experimental}.

Generalized linear models (GLM) extend linear models by applying a \emph{link} function that transforms the predicted distribution, e.g. $\log(Y) = aX + b \rightarrow \hat{Y} = \exp(aX + b)$ in the case of \emph{log link}. Generalized additive models (GAM) further extends GLM by applying transformations in each variable, e.g. $g(E[Y]) = f_1(x_1) + f_2(x_2) + ... + f_n(x_n) + b$, and discovering the most appropriate functions $f_i$ for each input variable \cite{hastie1990generalized}. The linear models are evaluated by Akaike Information Criterion (AIC), according to the maximum likelihood of estimates and number of parameters of the model \cite{akaike1974new}. We tested several linear models and settled our analysis with $DICE \sim ls+ls:Family+\log(Dataset) +\log(lr):ls:Family+\log(reg)$, which is presented in the next section along experiments, metrics, results and discussion.

\section{Experiments}
\label{sec:exps}

In this section we present the experimental configuration (\ref{sec:setup}), evaluation metrics (\ref{sec:metrics}), results (\ref{sec:results}) and discussion (\ref{sec:discussion}). In Section \ref{sec:results} (Results) we present the experimental results, report the DICE index and BCE Loss of all models, compare their differences, and analyse the model's dependences on the chosen parameters by linear regression models, ANOVA and HSD tests. In Section \ref{sec:discussion} (Discussion) we comment our findings in relation to literature and place general considerations.

\subsection{Experimental configuration}
\label{sec:setup}

For each of the 324 different combinations of variables (model, dataset, initial learning rate, L2 regularization), the model was trained from scratch using 5-fold cross-validation. The development dataset was divided into 5 equally-sized subsets (folds) and each one is used for validation while the remaining four are used to fit the model. For each fold, every combination of parameters is trained for 50 epochs or less. We stopped training when there was no improvement in the validation loss after 5 epochs (early stopping). This results in 1620 training/evaluation runs, corresponding to 5 folds × 324 different conditions.

After all runs, we computed the metrics and fitted a linear regression model to the experiment parameters and results, then performed statistical ANOVA and HSD tests to assess significant differences.
All experiments were performed in a Foxconn HPC M100-NHI with eight NVIDIA V100-SXM2-16GB GPUs, using the software we implemented with Python, PyDicom, Tensorflow, SegmentationModels, Slurm and StatsModels. Next we present the evaluation metrics, then results and discussion.

\subsection{Evaluation metrics}
\label{sec:metrics}

To evaluate the quality of each model, we use two metrics: BCE (Binary Cross-Entropy) loss and DICE index \cite{dice1945measures}. These are computed on the validation set of each training run, and also separately on the held-out test set from the corresponding dataset. We then compare all configurations across datasets and models, considering the best hyper-parameters for each configuration. BCE and DICE formulas are given below:
\begin{align*}
BCE &= - \frac{1}{N} \sum y \cdot \log(\hat{y}) + (1-y) \cdot \log(1-\hat{y})\\
DICE &= \frac{2TP}{2TP+FP+FN} = 2\frac{\sum y \cdot \lfloor\hat{y}\rceil}{\sum y + \lfloor\hat{y}\rceil}
\end{align*}

Where $TP, FP, FN$ are true positives, false positives and false negatives, respectively, $N$ is the number of pixels in the dataset, $y$ are the binary labels for all pixels, $\hat{y}$ are the corresponding predictions and $\lfloor \cdot \rceil$ denotes rounding. BCE is a positive value which gives the amount of entropy mismatch in the predictions, the lower the better. DICE range from 0\% (no correspondence) to 100\% (images are perfectly equal) and is equivalent to F1-score. In the next section we present the results of the experiments, our analyses and discussion.

\subsection{Results}
\label{sec:results}

\begin{figure}[htb]
    \centering
    \includegraphics[width=.49\columnwidth]{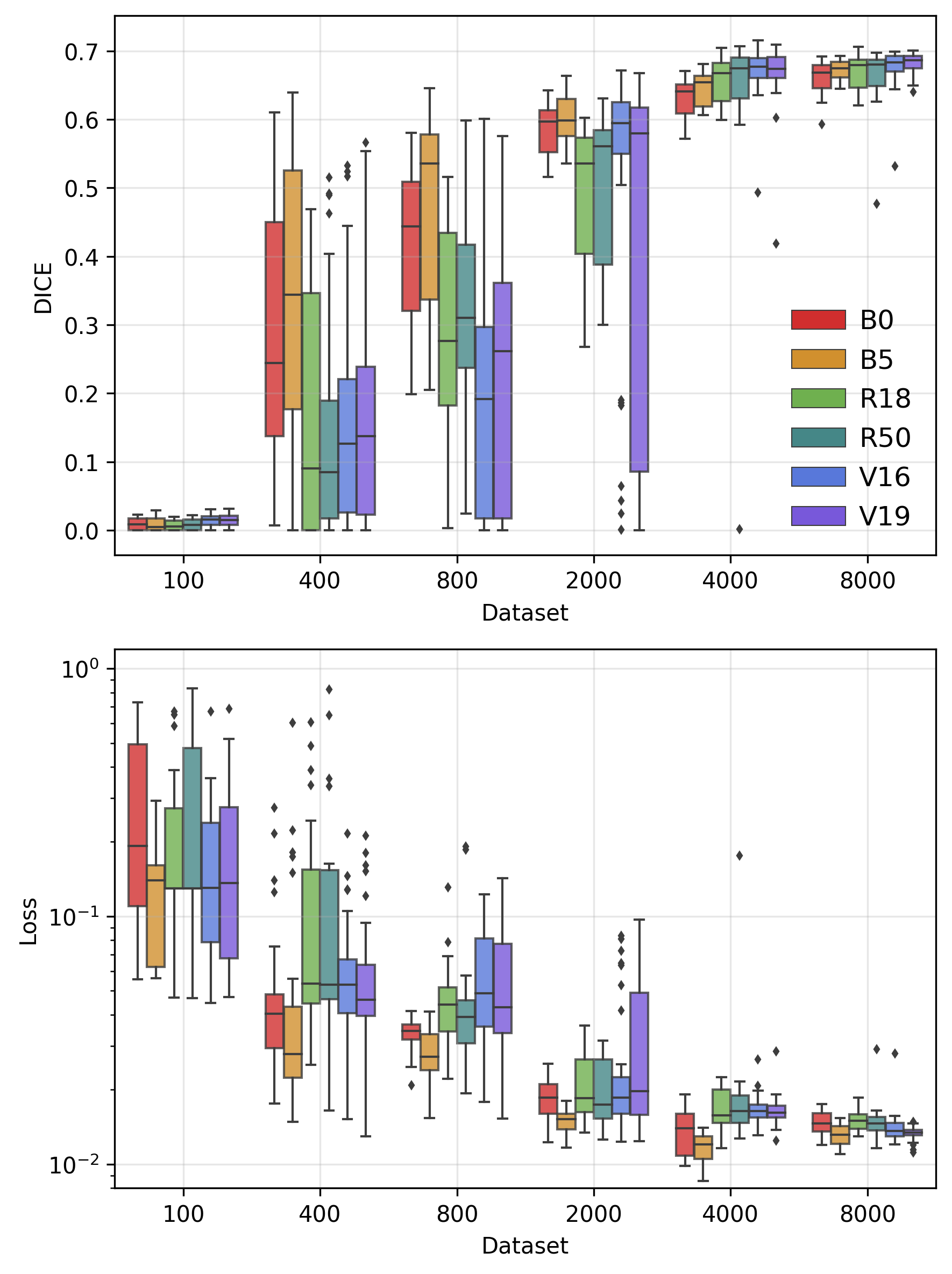}
    \includegraphics[width=.49\columnwidth]{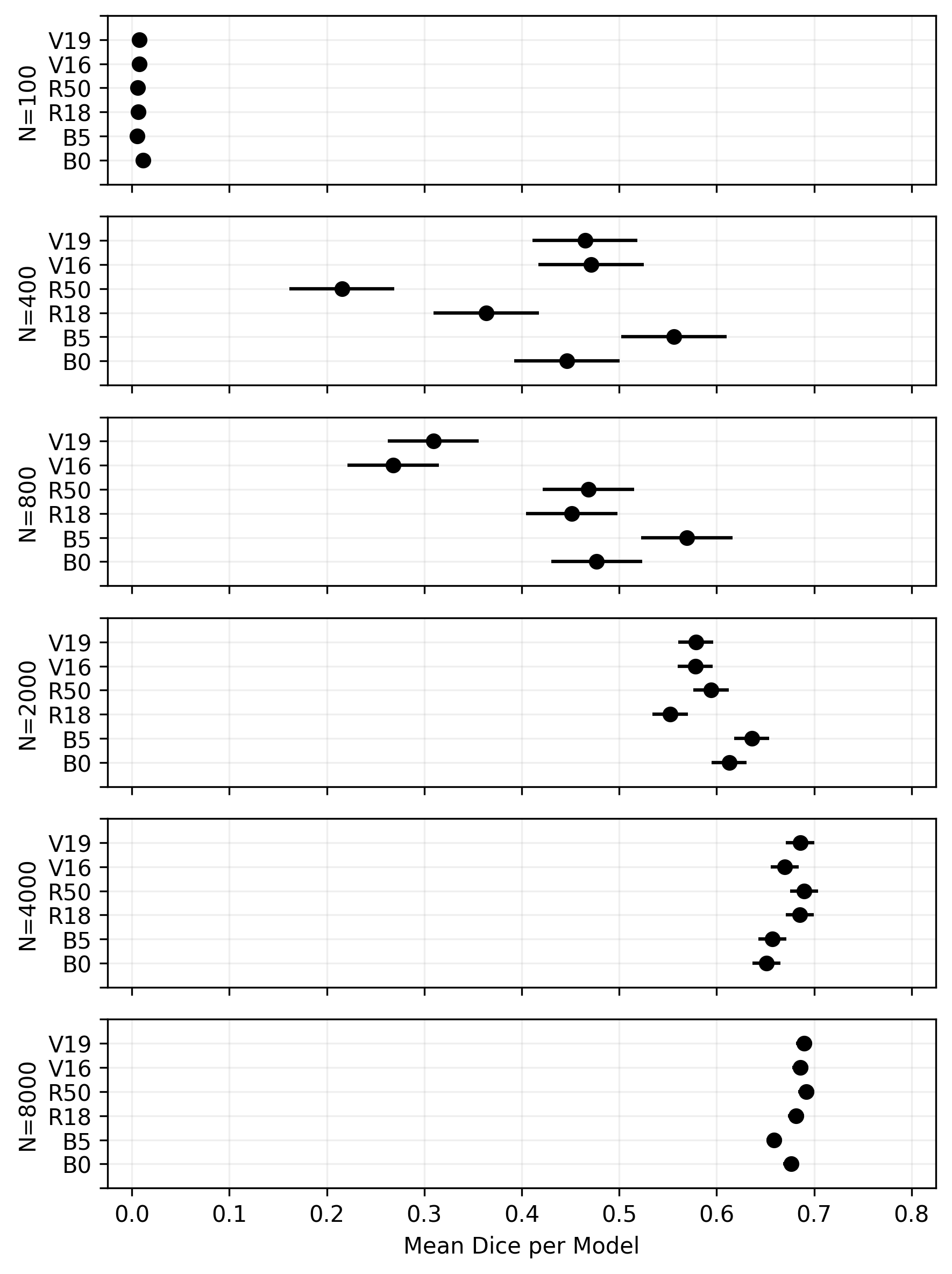}
    \caption{Evaluation of models: the left panel shows DICE index (upper) and BCE Loss (lower) on the test set for all 1,620 network models, in all datasets and all folds. Each dataset (x-axis) contains six medians and IQRs for B0, B5, R18, R50, V16 and V19, respectively, corresponding to EfficientNets B0 and B5, ResNets 18 and 50 and VGGs 16 and 19. All networks performed worse in the smaller datasets and similarly well on the larger datasets. We observe mixed results with a gradual performance increase in the middle ground between 400 and 4,000 images. The right panel shows Mean DICE scores across datasets and models. The error bars are 95\% confidence intervals calculated using HSD test. The results of all models are directly comparable, and overlapping intervals should be interpreted as the corresponding null hypothesis (no statistical difference).}
    \label{fig:eval}
\end{figure}

All networks performed badly on the smallest dataset (due to insufficient training cases) and improved gradually when fed with more data. The network performed similarly well on the larger datasets (Figure \ref{fig:eval}). However, the networks behaved very differently in the middle sizes (between 400 and 2000 samples), both in absolute performance and variability of the metrics. Loss function behaved similarly, showing a similar improvement related to increasing dataset size. Although, we opted to focus this report on DICE index because it is easier to interpret, has close relation to the task, and is common in literature.

\begin{figure}[htb]
    \centering
    \includegraphics[width=0.7\columnwidth]{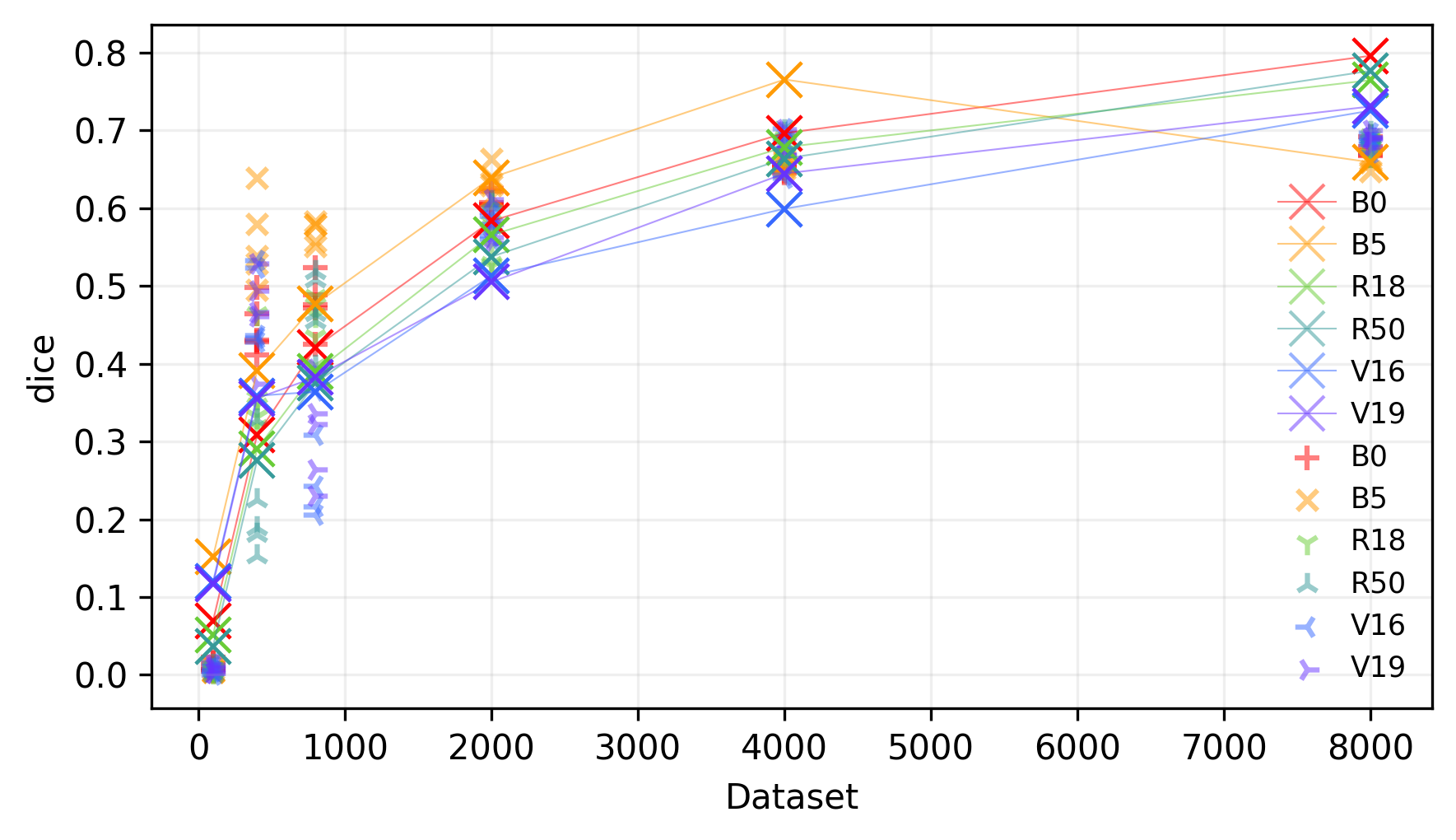}
    \caption{Predictions of DICE index for each model in each dataset size given by the linear regression model (large X marks connected with lines), and the observed values for the corresponding DICE data (small +, x and arrow marks).}
\label{fig:preds}
\end{figure}

After all runs, we fitted a linear regression model (Table \ref{tab:lm}) to the experiment parameters and results, then performed statistical ANOVA and HSD tests to assess significant differences (Table \ref{tab:anova} and Figure \ref{fig:eval}). The variables $Family, ls, lr$ and $reg$ correspond to the network architecture family ({V}GG, {R}esNet or {E}fficientNet), model version size ({L}ong or {S}hort, $ls$), initial learning rate ($lr$) and L2 regularization parameters ($reg$), respectively. The nesting of $Family$ within model version size ($ls/Family = ls + ls:Family$) in the regression formula is a way to access a rough effect of model size while not treating different architectures (long and short versions) as equivalents.

We also fitted a linear model with a similar formula and using a log link function instead of the log transformations on predictors: $\log(DICE) \sim ls + ls:Family + Dataset + lr:ls:Family + reg$, but it had a bigger AIC value of -582. According to the results, the dataset size had overall larger absolute coefficients (Table \ref{tab:lm}) and more significant effect (Tables \ref{tab:lm} and \ref{tab:anova}) in DICE than architecture, initial learning rates or L2 regularization values in the linear model. HSD test was applied to the mean DICE and confidence intervals (Figure \ref{fig:eval}). EfficientNets had consistently higher DICE scores up to 2000 images. ResNets and VGGs required more data to achieve a comparable level of performance, but slightly surpassed EfficientNets in the two largest datasets. In the next section we discuss these results in relation to other evidence in literature, and place general considerations.

\begin{table}[htb]
    \centering{\sffamily\footnotesize
    \begin{tabular}{lrrc}
    \hline
    \textbf{Term} & \textbf{coef.} &\textbf{z} &\textbf{p-value} \\
    \hline
    Intercept &-0.5810 &-178.383 &0.0000\\
    $ls[T.short]$ &0.0016 &0.039 &0.9686\\
    $ls[long]:Family[T.ResNet]$ &-0.0936 &-23.476 &0.0189\\
    $ls[short]:Family[T.ResNet]$ &-0.1107 & -27.771 &0.0055\\
    $ls[long]:Family[T.VGG]$ &-0.3903 &-97.863 &0.0000\\
    $ls[short]:Family[T.VGG]$ &-0.3237 &-81.173 &0.0000\\
    $\log(Dataset)$ &0.1627 &787.068 &0.0000\\
    $\log(lr):ls[long]:Family$ &0.0158 &40.202 &0.0001\\
    $\log(lr):ls[short]:Family[EfficientNet]$ &0.0193 &49.085 &0.0000\\
    $\log(lr):ls[long]:Family[ResNet]$ &0.0130 &32.955 &0.0010\\
    $\log(lr):ls[short]:Family[ResNet]$ &0.0111 &28.156 &0.0049\\
    $\log(lr):ls[long]:Family[VGG]$ &-0.0281 &-71.467 &0.0000\\
    $\log(lr):ls[short]:Family[VGG]$ &-0.0193 &-49.051 &0.0000\\
    $\log(reg)$ &-0.0003 &-0.383 &0.7021\\
    \hline
    \end{tabular}
    \\
    {\small Model: $DICE \sim ls + ls:Family + log(Dataset) + log(lr):ls:Family + log(reg)$}}
    \caption{Coefficients and p-values for the linear model, with AIC: -2213.}
    \label{tab:lm}
\end{table}

\begin{table}[htb]
    \centering{\sffamily\footnotesize
    \begin{tabular}{lrrrc}
    \hline
    \textbf{Term} &\textbf{SS} &\textbf{F} &\textbf{p-value} &$\eta^2_p$\\
    \hline
    $ls$ &0.0145 &0.9802 &0.3223 &0.0001\\
    $ls:Family$ &1.7284 &29.1844 &<10$^{-22}$ &0.0145\\
    $\log(Dataset)$ &91.7177 &6194.7601 &<10$^{-30}$ &0.7693\\
    $\log(lr):ls:Family$ &1.9866 &22.3631 &<10$^{-24}$ &0.0167\\
    $\log(reg)$ &0.0022 &0.1463 &0.7021 &0.0000\\
    \hline \hline
    Residuals &23.7779 &- &- &0.1994\\
    \hline
    \end{tabular}}
    \caption{ANOVA table for linear model variables and their interaction. $\eta^2_p$ is the effect size, measuring the proportion of SS explained by each term, while $Residuals$ measure the unknown SS. The terms $ls:Family$, $\log(Dataset)$ and $\log(lr):ls:Family$ are responsible for 80\% of DICE index variance in the trained models. We observe that gross performance increment comes from larger datasets, while learning rates and network size give small but significant increments depending on the network architecture ($Family$).}
    \label{tab:anova}
\end{table}

\subsection{Discussion}
\label{sec:discussion}


Apparently, smaller models (EfficientNet-B0, VGG16 and ResNet18) are not too small for LV segmentation tasks, and were able to achieve good levels of performance when given sufficient data. One could expect that different models would reach different plateaus. But we observed in all families that, as the training dataset size increases, the gain in DICE performance of deeper networks vanishes as the performance of shallower models levels up. Considering each network family separately, the network size effect is greater for the EfficientNets than for the ResNets and VGGs (Table \ref{tab:lm} coefficients).
As in the case of ImageNet classification, the gain of scaling up is bigger for the EfficientNets than in other families. We noticed that in all dataset sizes B5 performed significantly better than B0 (Figure \ref{tab:anova}), showing that it did not over-fit more easily with the increased number of parameters, even considering that B0 reached comparable performance with more data.

These results contradict the common sense that bigger models would suffer more from over-fitting in smaller datasets, decreasing their performance, and that they would gain more performance as more data was made available \cite{martin2018rethinking}. Our findings with EfficientNet support the idea of \citet{neyshabur2019towards} that bigger models does not necessarily over-fit, but this pattern does not hold for other families. Considering the ResNets, the deeper ResNet50 starts with a DICE performance below the shallow counterpart (ResNet18), but improved faster and caught up with 4000 samples. We observed that VGGs never differed significantly, probably because of their smaller differences in depth and number of parameters, and they also have high sensitivity to hyper-parameters.
VGG16 had a particular tipping point from the worst results in the 800 images dataset to scoring near the top contenders on the next dataset with 2000 images.

A general consideration in the intermediate dataset sizes is that all models were relatively more sensitive to hyper-parameters than to architecture, as their performances had higher variance than when more data was made available. This might affect sample size and model choice, as with 2000 images EfficientNet B5 performed significantly better than the ones from other families, but with 8000 images all models had no practical difference.
Thus, in scenarios where bigger datasets are unavailable and acquisition is costly, it might be possible to use smaller data samples with deeper networks and careful optimization. When a bigger dataset size is available, neither model selection or optimization provides much performance gain, different approaches such as changing pre-processing or loss function might be required. However, as in the demonstrated case, any method should be tested in a range of training dataset sizes in order to guarantee its performance. In the next section we conclude this paper, summarizing our goals, experiments, results and future investigations.

\section{Conclusion}
\label{sec:conclusion}

In this work we described a framework to check when the choice for the network to be used in an application might change due to the number of images available in a training dataset. We demonstrated an application of this framework for LV segmentation in cardiac MRI images using the LSVC dataset. In our experiments we observed the relationship between prediction error, dataset size and network size; and also identified hyper-parameter effects on network performance, by modelling with linear model $DICE \sim ls+ls:Family+\log(Dataset)+\log(lr) :ls:Family+\log(reg)$ and executing and ANOVA of its interaction terms.
The results indicated that: sample size affected performance more than architecture or hyper-parameters; in small samples the performance was more sensitive to hyper-parameters than architecture; the performance difference between shallow and deeper networks was not the same across families.

Further investigation and extension of these experiments will focus on analysing the regularization effects' patterns regarding network size. Also, we plan to use similar strategies in Neural Architecture Search (NAS) and Automatic Deep Learning (AutoDL) to discover better segmentation networks for evolving and multi-centric medical imaging datasets.

\section*{Acknowledgement}
We thank Zerbini Foundation and FoxConn for supporting this work; A. Marco, C. Graves, D. Cardenas, J. R. Ferreira Jr. and R. F. Pereira for discussing ideas about experiments; R. Moreno and M. Rebelo for initial manuscript revision.

\bibliographystyle{abbrvnat}
\bibliography{refs}
\end{document}